\documentclass[10pt,twocolumn,letterpaper,twocolumn]{article}
\usepackage{ol2}
\usepackage[draft]{hyperref}
\usepackage{amsmath}
\usepackage{calc}
\usepackage{amsfonts}
\usepackage{amssymb}
\usepackage{graphicx}
\usepackage{bm}
\usepackage{color}
\usepackage{epsfig}
\usepackage{ifthen}

\def\be{\begin{equation}}
\def\ee{\end{equation}}
\def\ignore#1{}

\begin{document}

\twocolumn[
\title{An achromatic polarization retarder realized with slowly varying linear and circular birefringence}

\author{A. A. Rangelov}

\address{
Department of Physics, Sofia University, James Bourchier 5 blvd.,
1164 Sofia, Bulgaria }

\begin{abstract}
Using the phenomena of linear and circular birefringence we
propose a device that can alter general elliptical polarization of
a beam by a predetermined amount, thereby allowing conversion
between linearly-polarized light and circularly polarized light or
changes to the handedness of the polarization. Based on an analogy
with two-state adiabatic following of quantum optics, the proposed
device is insensitive to the frequency of the light -- it serves
as an achromatic polarization retarder.
\end{abstract}


 ] 

Birefringence, the separation of a ray of light into two rays when
it passes through some anisotropic materials such as quartz,
 was first described in 1669 by Bartholinus, who observed it in
calcite \cite{Wolf,Azzam,Theocaris,Goldstein,Saleh}. Some 30 years
later Huygens explained the phenomenon of birefringence by
suggesting different refractive indices for ordinary and
extraordinary rays. Another type of birefringence is the optical
activity; as light passes through material in which the left- and
right-circularly polarized components travel with different phase
velocities \cite{Wolf,Azzam,Theocaris,Goldstein,Saleh} the plane
of linearly polarized light rotates, a phenomenon also known as
circular birefringence.

Natural optical activity was first observed in quartz by Arago in
1811. In 1845 Faraday discovered that a similar phenomenon could
be induced by a static magnetic field: when linearly polarized
light propagated in a dielectric medium parallel to a strong
static magnetic field the plane of polarization rotated. This
artificial optical activity is now known as the Faraday effect \cite%
{Wolf,Azzam,Theocaris,Goldstein,Saleh}.

The theory underlying changes to optical polarization as a beam of light passes through matter relies on two components of the electric field, often parametrized by Stokes parameters
\cite{Wolf}
 or by a point on the Poincar\'e sphere. For present purposes the Jones vector
 \cite{Jon41}
 provides a more convenient tool, because it offers a simple analogy with the complex-valued probability amplitudes of the traditional two-state atom.
As will be shown, such a treatment allows the design of a device
that will convert any input polarization state into a prescribed
polarization state, independent of the wavelength of the light.
The proposed technique is analogous to that of adiabatic-following
in quantum optics \cite{Allen,Sho90,Vit01a,Vit01b,Sho08} and has
the same advantages of efficiency and robustness.

A numerical investigation of broadband conversion from circularly
polarized light into linearly polarized light for the wavelength
range of 434 nm to 760.8 nm for crystalline quartz was made by
Darsht et al. \cite{Darsht}. This result is a special case of the
general adiabatic technique described here.

The basic equation that describe the propagation of plane waves
through an optically active uniaxial linear crystal with
negligible absorption (quartz for example) is the wave equation
for the electric field $\mathbf{E}$,
 \begin{equation}
\nabla ^{2}\mathbf{E=}\frac{\mathbf{\epsilon }}{c^{2}}\cdot
\frac{\partial ^{2}\mathbf{E}}{\partial t^{2}}. \label{wave
equation}
\end{equation}%
where $\mathbf{\epsilon }$ is the electric permittivity tensor and
$c$ is the light velocity in vacuum.

 We shall deal with light propagation in one dimension. Let a monochromatic
plane wave with frequency $\omega $ propagate through the inhomogeneous
crystal along the $z$ axis:
\begin{equation}
\mathbf{E}\left( z,t\right) \mathbf{=}\left[
\begin{array}{c}
E_{x}\left( z\right) \mathbf{e}^{\left( ikz-\omega t\right) } \\
E_{y}\left( z\right) \mathbf{e}^{\left( ikz-\omega t\right) } \\
0%
\end{array}%
\right] .
\end{equation}%

Let the propagation axis be $z$, along one of the principal axes
of the crystal, and let the other two optical axes be $x$ and $y$.
In this reference frame the dielectric tensor $\mathbf{\epsilon }$
takes the form
\cite%
{Goldstein,Saleh}
\begin{equation}
\mathbf{\epsilon }=\left[
\begin{array}{ccc}
n_{e}^{2} & iG & 0 \\
-iG & n_{o}^{2} & 0 \\
0 & 0 & n_{o}^{2}%
\end{array}%
\right] , \label{electric permittivity tensor}
\end{equation}%
where $n_{o}$ and $n_{e}$ are the ordinary and extraordinary
refractive indices respectively. The off-diagonal elements of this
tensor, denoted $G$, are responsible for the optical activity.

Because the electric field $\mathbf{E}\left( z,t\right) $ depends
only on the longitudinal coordinate $z$, we can replace $\nabla
^{2}$\ by $\partial ^{2}/\partial z^{2}$. The equation for $E_{z}$
is trivial and will not be considered further. The equations for
the transverse components of the field are the two scalar
equations
\begin{subequations}
\label{two scalar equations}
\be
\frac{\partial ^{2}E_{x}}{\partial z^{2}}
+2ik\frac{\partial E_{x}}{\partial z%
}-k^{2}E_{x}
=
-\frac{\omega ^{2}}{c^{2}}\left(
n_{e}^{2}E_{x}+iGE_{y}\right) ,
\ee
\be
\frac{\partial ^{2}E_{y}}{\partial z^{2}}+2ik\frac{\partial E_{y}}{\partial z%
}-k^{2}E_{y} =
\frac{\omega ^{2}}{c^{2}}\left(
i GE_{x}-n_{o}^{2}E_{y}\right)
\ee%
\end{subequations}
where
\begin{equation}
k^{2}=\frac{\omega ^{2}n_{o}^{2}}{c^{2}}.
\end{equation}%
Let us neglect any abrupt change of dielectric properties, such as occurs at interfaces (these are responsible for reflection) and consider only slowly varying fields,
for which
\begin{equation}
\left\vert \frac{\partial ^{2}E_{x,y}}{\partial z^{2}}\right\vert \ll
\left\vert 2k\frac{\partial E_{x,y}}{\partial z}\right\vert .
\end{equation}%
This condition requires that, over a distance comparable to the
optical wavelength, the fractional change of the field amplitude
should be much smaller than unity. The resulting {\em
slowly-varying amplitude approximation} \cite{Boyd} leads to the
equations
\begin{equation}
i\frac{\partial }{\partial z}\left[
\begin{array}{c}
E_{x} \\
E_{y}%
\end{array}%
\right] =\frac{\omega }{2cn_{o}}\left[
\begin{array}{cc}
n_{e}^{2}-n_{o}^{2} & -iG \\
iG & 0%
\end{array}%
\right] \left[
\begin{array}{c}
E_{x} \\
E_{y}%
\end{array}%
\right] , \label{first amplitude equation}
\end{equation}%
These are the equations upon which the present proposal is based.

The two complex-valued field amplitudes $E_x$ and $E_y$ are elements of a
Jones vector ${\bf J}$
\cite{Jon41}.
We symmetrize the
 the diagonal terms $n_{e}^{2}-n_{o}^{2}$ and $0$ in Eq. (\ref%
{first amplitude equation}) by incorporating an identical phase in
the amplitudes $E_{x}$ and $E_{y}$. Upon making this choice we
have \ignore{
\begin{equation}
i\frac{\partial }{\partial z}\left[
\begin{array}{c}
E_{x} \\
E_{y}%
\end{array}%
\right] =\frac{\omega }{4cn_{o}}\left[
\begin{array}{cc}
n_{e}^{2}-n_{o}^{2} & -2iG \\
2iG & n_{o}^{2}-n_{e}^{2}%
\end{array}%
\right] \left[
\begin{array}{c}
E_{x} \\
E_{y}%
\end{array}%
\right] .
\end{equation}%
Now we can rewrite the last equation as
}
\begin{equation}
i\frac{\partial }{\partial z}\left[
\begin{array}{c}
E_{x} \\
E_{y}%
\end{array}%
\right] =\frac{1}{2}\left[
\begin{array}{cc}
-\Delta & -i\Omega \\
i\Omega & \Delta%
\end{array}%
\right] \left[
\begin{array}{c}
E_{x} \\
E_{y}%
\end{array}%
\right] , \label{two-state atom}
\end{equation}%
where
\begin{subequations}
\label{coupling and detuning}
\begin{eqnarray}
\Omega &=&\frac{\omega G}{cn_{o}}, \label{Rabi frequency} \\
\Delta &=&\frac{\omega \left( n_{o}^{2}-n_{e}^{2}\right) }{2cn_{o}}.
\label{Detuning}
\end{eqnarray}
\end{subequations}
As a final step we map the coordinate dependance into time
dependance, $z = c t$. By so doing Eq. (\ref{two-state atom})
becomes the traditional time-dependent Schr\"{o}dinger equation
for a two-state atom in the rotating-wave approximation
\cite{Allen,Sho90,Vit01a,Vit01b,Sho08}. The two field amplitudes
$E_{x}$ and $E_{y}$ are analogs of the probability amplitudes for
the ground state (horizontal polarization) and the excited state
(vertical polarization). The off-diagonal element $\Omega $ in Eq.
(\ref{two-state atom}) is known as Rabi frequency, while the
element $\Delta $ corresponds to the atom-laser detuning
\cite{Allen,Sho90}.

Let us assume that $\Omega $ and $\Delta $ are function of the coordinate $z$%
. Then we can write Eq. (\ref{two-state atom}) in the so-called adiabatic
basis \cite{Allen,Sho90,Vit01a,Vit01b,Sho08} (for the two-state atom this is
the basis of the instantaneous eigenstates of the Hamiltonian):
\begin{equation}
i\frac{\partial }{\partial z}\left[
\begin{array}{c}
E_{x}^{A} \\
E_{y}^{A}%
\end{array}%
\right] =\left[
\begin{array}{cc}
-\frac{1}{2}\sqrt{\Omega ^{2}+\Delta ^{2}} & \partial \varphi / \partial z%
  \\
\partial \varphi / \partial z & \frac{1}{2}\sqrt{\Omega ^{2}+\Delta
^{2}}%
\end{array}%
\right] \left[
\begin{array}{c}
E_{x}^{A} \\
E_{y}^{A}%
\end{array}%
\right] ,  \label{adiabatic equation}
\end{equation}%
The connection between the Jones vector $\mathbf{J}\left( z\right) =\left(
E_{x},E_{y}\right) $ in the original basis and the Jones vector $\mathbf{J}%
^{A}\left( z\right) =\left( E_{x}^{A},E_{y}^{A}\right) $ in the adiabatic
basis is given by
\begin{equation}
\mathbf{J}\left( z\right) =R\left( z\right) \mathbf{J}^{A}\left( z\right) ,
\end{equation}%
where $R\left( z\right) $ is the involutory matrix
\begin{equation}
R\left( z\right) =\left[
\begin{array}{cc}
\cos \left( 2\varphi \right) & i\sin \left( 2\varphi \right) \\
-i\sin \left( 2\varphi \right) & -\cos \left( 2\varphi \right)%
\end{array}%
\right] ,
\end{equation}%
with angle $\varphi$ defined by the equation%
\begin{equation}
\tan \left( 2\varphi \right) =\frac{\Omega }{\Delta }.  \label{angle}
\end{equation}

For adiabatic evolution of the system there are no transitions
between the amplitudes $E_{x}^{A}$ and $E_{y}^{A}$. Hence
$\left\vert E_{x}^{A}\right\vert $ and $\left\vert
E_{y}^{A}\right\vert $ remain constant
\cite{Allen,Sho90,Vit01a,Vit01b,Sho08}. Mathematically, adiabatic
evolution means that in Eq. (\ref{adiabatic equation}) the
non-diagonal terms can be neglected compared to the diagonal
terms. This occurs when \cite{Allen,Sho90,Vit01a,Vit01b,Sho08}
\begin{equation}
\left\vert \frac{\partial \varphi }{\partial z}\right\vert \ll \left\vert
\sqrt{\Omega ^{2}+\Delta ^{2}}\right\vert .  \label{adiabatic condition}
\end{equation}%
Thus adiabatic evolution requires smooth $z$-dependance of $\Omega $, $%
\Delta $ and large $\sqrt{\Omega ^{2}+\Delta ^{2}}$. For a pure adiabatic
evolution, the evolution matrix in adiabatic basis is diagonal and contains
only phase factors:%
\begin{equation}
U^{A}\left( z_{f},z_{i}\right) =\left[
\begin{array}{cc}
\exp \left( i\eta \right) & 0 \\
0 & \exp \left( -i\eta \right)%
\end{array}%
\right] ,
\end{equation}%
with the adiabatic phase
\begin{equation}
\eta =\int_{z_{i}}^{z_{f}}\frac{1}{2}\sqrt{\Omega ^{2}+\Delta ^{2}}dz.
\end{equation}%
The evolution matrix in the original basis (the Jones matrix) is%
\begin{equation}
U\left( z_{f},z_{i}\right) =R\left( z_{f}\right) U^{A}\left(
z_{f},z_{i}\right) R\left( z_{i}\right) ,
\end{equation}%
or explicitly:
\begin{subequations}
\label{adiabatic solution}
\begin{eqnarray}
U_{11}\left( z_{f},z_{i}\right) &=&e^{i\eta }\cos \varphi \left(
z_{i}\right) \cos \varphi \left( z_{f}\right)  \notag \\
&&+e^{-i\eta }\sin \varphi \left( z_{i}\right) \sin \varphi \left(
z_{f}\right) , \\
U_{12}\left( z_{f},z_{i}\right) &=&ie^{i\eta }\sin \varphi \left(
z_{i}\right) \cos \varphi \left( z_{f}\right)  \notag \\
&&-ie^{-i\eta }\cos \varphi \left( z_{i}\right) \sin \varphi \left(
z_{f}\right) , \\
U_{21}\left( z_{f},z_{i}\right) &=&ie^{-i\eta }\sin \varphi \left(
z_{i}\right) \cos \varphi \left( z_{f}\right)  \notag \\
&&-ie^{i\eta }\cos \varphi \left( z_{i}\right) \sin \varphi \left(
z_{f}\right) , \\
U_{22}\left( z_{f},z_{i}\right) &=&e^{-i\eta }\cos \varphi \left(
z_{i}\right) \cos \varphi \left( z_{f}\right)  \notag \\
&&+e^{i\eta }\sin \varphi \left( z_{i}\right) \sin \varphi \left(
z_{f}\right) .
\end{eqnarray}
\end{subequations}
Equation (\ref{adiabatic solution}) gives the general adiabatic
evolution scenario for an arbitrary polarization.

\emph{\textbf{\ Achromatic conversion:}} When initially the light
is linearly polarized in the horizontal direction,
$\mathbf{J}\left( z_{i}\right) =\left( 1,0\right) $, and if we
ensure the initial condition $\varphi \left( z_{i}\right) =\pi
/2$, then from Eq. (\ref{adiabatic solution}) we have
\begin{subequations}
\begin{eqnarray}
J_{1}\left( z_{f}\right)  &=&e^{-i\eta }\sin \varphi \left( z_{f}\right) , \\
J_{2}\left( z_{f}\right)  &=&ie^{-i\eta }\cos \varphi \left( z_{f}\right) .
\end{eqnarray}%
\end{subequations}
Obviously from the last equation the global phase $\eta $ is
unimportant and can be removed from final Jones vector components,
therefore such polarization conversion will be frequency
independent. For such achromatic conversion we can end up with
left
circularly polarized light, $\mathbf{J}\left( z_{f}\right) =\left( 1/\sqrt{2}%
,i/\sqrt{2}\right) $, if we set the final angle $\varphi (z_{f})=\pi /4$.
This process is reversible: if we start with left circularly polarized light
with initial angle $\varphi (z_{i})=\pi /4$ and we finished with angle $%
\varphi (z_{f})=\pi /2$, then we achieve reversal of the direction of motion
and we end up with a horizontal polarization.

Analogously, if we set the final angle to be $\varphi (z_{f})=-\pi /4$, then
we end up with right circularly polarized light, $\mathbf{J}\left(
z_{f}\right) =\left( 1/\sqrt{2},-i/\sqrt{2}\right) $, instead of left
circularly polarized light. Again, the process is reversible.

Alternatively we can start with horizontal polarized light and applying the
conditions $\varphi \left( z_{i}\right) =\pi /2$, $\varphi \left(
z_{f}\right) =0$, then we will end in vertical polarized light.

Finally combining two or more then two achromatic conversion from
the above processes we can have achromatic conversion from left
circularly polarized light to right circularly polarized light and
vice versa, or from left (right) circularly polarized light to
vertical polarized light and vice versa.

The proposed achromatic retarder
can be realized in an optically active uniaxial linear crystal such as quartz,
which in addition exhibits both stress-induced linear birefringence and
circular birefringence from the Faraday effect. \newline
From the definition of the angle $\varphi $ (Eq.(\ref{angle})), we see that
the needed values of $0,\pm \pi /4$ and $\pi /2$ would be achieved when:
\begin{subequations}
\begin{eqnarray}
&&\varphi \underset{\Omega /\Delta \rightarrow 0+}{\rightarrow }0, \\
&&\varphi \underset{\Omega /\Delta \rightarrow \pm \infty }{\rightarrow }\pm
\pi /4, \\
&&\varphi \underset{\Omega /\Delta \rightarrow 0-}{\rightarrow }\pi /2.
\end{eqnarray}
\end{subequations}
The proposed adiabatic retarder has the useful property that the
polarization conversion depends only on the initial and final
values of angle $\varphi $. The retarder is frequency independent
and it is robust against variations of the propagation length,
rotary power, etc., in contrast to the traditional retarders. The
only restriction that we have to follow during
the polarization conversion is the adiabatic evolution condition (see Eq.(%
\ref{adiabatic condition})), thus slow change of angle $\varphi $.

In conclusion, we have used the analogy between the equation that
describes the polarization state of light propagating through an
optically active anisotropic medium, and the time-dependent
Schr\"{o}dinger equation applied to adiabatic change of a two
state atom, to propose effective and achromatic polarization
retarder.

This work has been supported by the European Commission project
FASTQUAST, the Bulgarian NSF grants D002-90/08, DMU02-19/09,
IRC-CoSiM and Sofia University Grant 022/2011. The author is
grateful to Nikolay Vitanov for stimulating discussions and to
Bruce Shore for critical reading of the manuscript.

\end{document}